**Nerve Impulses Have Three Interdependent Functions : Communication, Modulation And Computation.**

William Winlow[1,2] and Andrew S. Johnson[1]

[1]Dipartimento di Biologia, Università degli Studi di Napoli, Federico II, Via Cintia 26, 80126 Napoli, Italia; [2]Institute of Ageing and Chronic Diseases, The Apex Building, West Derby Street, University of Liverpool, Liverpool, L7 8TX, UK

> *"The great sins of the world take place in the brain, but it is in the brain that everything takes place.... it is in the brain that the poppy is red, that the apple is odorous, that the skylark sings"* (Wilde, 1891 [105]) and the soliton and the action potential are the primary elements underlying sentience [5] – they need to be better understood.

## Abstract

Comprehending the nature of action potentials is fundamental to our understanding of the functioning of nervous systems in general. Here we consider their evolution and describe their functions of communication, modulation and computation within nervous systems. The ionic mechanisms underlying action potentials in the squid giant axon were first described by Hodgkin and Huxley in 1952 and their findings have formed our orthodox view of how the physiological action potential functions. However, substantial evidence has now accumulated to show that the action potential is accompanied by a synchronized coupled soliton pressure pulse in the cell membrane, the action potential pulse (APPulse). Here we explore the interactions between the soliton and the ionic mechanisms known to be associated with the action potential. Computational models of the action potential usually describe it as a binary event, but we suggest that it is quantum ternary event known as the computational action potential (CAP), whose temporal fixed point is threshold, rather than the rather plastic action potential peak used in other models. The CAP accompanies the APPulse and the Physiological action potential. Therefore, we conclude that nerve impulses appear to be an ensemble of three inseparable, interdependent, concurrent states: the physiological action potential, the APPulse and the CAP.

## INTRODUCTION

Traditionally a nerve impulse has been considered to be an electrochemical phenomenon with experiments dating back 250 years to Galvani [1,2]. However, this assumption has prevented contemporary consideration of the issues surrounding computation and assumes the temporal and communicative aspects of nervous activity can be resolved by electrical theory. In other words, much has been done to understand the biophysical mechanisms underlying nerve action potentials, as a result of the excellent work by Hodgkin and Huxley [3] and their successors. However, substantial findings that a soliton wave accompanies the physiological action potential [4] is not yet widely accepted, nor are suggestions that a computational form of the action potential [5] may also exist. Although the Hodgkin-Huxley model describes the ionic movements that lead to changes in electrical potentials across nerve cell membranes, which lead to action potentials, it cannot by itself provide the temporal accuracy of computation. However, when combined with the soliton to form the action potential pulse [5] (APPulse), the ionic movements associated with the action potential may be seen as part of a process that provides entropy to the soliton. This then defines the speed of the pulse which determines the temporal characteristics that define the computational characteristics of each neuron. Here we explore these concepts in more detail and consider whether they are interrelated manifestations of the nerve impulse. To set the scene we start by considering the evolution of action potentials and then their functions within nervous systems.

## Evolution of Action Potentials

If we are to understand the nature of neuronal action potentials, we first need to understand how they might have evolved. However, although action potentials occur in many neurons, they are not by themselves a diagnostic feature of neurons as we will see below because they also occur in a wide variety of tissues and organisms, including plant cells and bacteria.

*Phylogenetic considerations* In the nerveless placozoa, the simplest known free-living animals, inducible fast sodium action potentials have been demonstrated, [6] as have $Ca_v3$-like channels. This demonstrates that electrical signalling occurs ubiquitously in cells other than neurons or muscle, including unicellular organisms [7], plants [8] and may also occur in some human carcinomas where electrical activity appears to be driven by voltage-gated sodium channels [9,10]. Electrical signalling also occurs in bacterial biofilm communities through release of intracellular potassium [11,12] which depolarizes nearby cells via voltage gated potassium channels ($K_V$) [13] and coordinates metabolic states among the cells.[11] This depolarizing wave occurs in the absence of voltage-gated sodium ($Na_v$) or calcium channels ($Ca_v$), which are commonly found in bilaterian nervous systems [14] alongside $K_v$ channels. Additionally, action potentials have also evolved in fungi and plant cells, [8] but are much slower than those of the squid giant axon, by a factor of $10^3$ and depolarisation is driven by efflux of chloride ions and influx of $Ca^{2+}$, which opens the $Cl^-$ channels, rather than influx of $Na^+$.[15] Repolarisation appears to be driven by increased permeability to potassium and protons. [15,16]

*Which membrane channels came first?* Kristan[14] speculates that $K_v$ channels, used to regulate cell volume in response to membrane stretch, were the only voltage-gated channels in the earliest animals, followed by $Ca_v$ channels to control metabolic state, regulate cilia [7] and muscle contractions. However, given that calcium is highly toxic to cytoplasm, calcium channels may have been the first to evolve alongside appropriately linked metabolic pumps for calcium removal. A combination of these channels could generate relatively slow action potentials in small soft-bodied animals at the dawn of metazoan evolution, so what drove the addition of $Na_v$ channels? Was it the need for more rapid movements as predation of one species on another became prevalent? We will never know, but whatever was the original selective advantage, shorter duration action potentials evolved, enhanced by the probable parallel evolution of fast $K_v$ channels to rapidly terminate them. Furthermore, Kristan [14] also provides evidence that neurons may have evolved more than once by parallel evolution, most usually from epidermal cells, [17] but in some cases from endodermal cells, e.g., in cnidaria, [18] as also appears to be true in gastropod molluscs such as *Aplysia.* [19] This suggests that wherever the appropriate genetic machinery is available to generate the appropriate channel proteins, action potentials may result. This appears to be the case in several human carcinoma models, where voltage gated sodium channels [9,10,20] and calcium channels [21] have been described and may be involved in action potential generation.

From an evolutionary standpoint it might be best to consider neurons as stretched secretory cells [19, 22], sometimes modified to communicate over long distances by means of fast action potentials, as in the case of alpha motor neurons. In other cases, the cells remain short and transmit by graded transmission, as in the case of the spikeless neurons of the retina or the olfactory bulb in vertebrates [23, 24] and many other examples in invertebrates. [25] Thus, there is no such thing as an orthodox neuron: they have hugely variable morphologies and nerve cell membranes vary in their properties, depending on their location and functions. However, once the appropriate genetic machinery evolved, action potentials became possible and are utilized where necessary in nervous systems where they are known to serve several functions.

**Neuronal action potentials have multiple functions**

Action potentials in neurons serve a number of functions in the nervous system, which may be summarised into the following three areas:

- *Communication*:
  - wiring the nervous, sensory effector and neurosecretory systems during development [26]
  - transmission within the adult nervous system, including formation and maintenance of synaptic connections [27,28,29]
- *Modulation* of synaptic function and memory storage after learning [30,31]
- *Computation* within neurons and brain neural networks [5,32,33].

**COMMUNICATION WITHIN NERVOUS SYSTEMS**

Action potentials can be conducted quite rapidly in invertebrate and vertebrate myelinated axons [34] (up to about 120 m/s in alpha motor neurons) and large diameter invertebrate fibres such as the squid giant axons (ca. 25 m/s), each of which are in essence a fusion of the axons of two interneurons. [35] However, this does not appear to be rapid enough for the process of neurocomputation (see below). In addition, there is a timing problem when the action potential peak is used as a time point as in many computational models (e.g., [36]). This is because action potentials are plastic phenomena [37,38] and the timing of the action potential peak varies considerably with spike frequency making it unsuitable for computational modelling and coding in the brain. This is demonstrated in Figure 1, where successive action potential peaks vary temporally from one another (Fig 1b). However, the threshold point of initiation (Fig1c) remains constant. Thus, threshold is a more appropriate temporal fixed point for neural communication and subsequent computation.[33,39] In non-spiking neurons, the threshold will be the point at which voltage-gated channels open to instigate transmitter release.

**There is accumulating evidence that both electrical and mechanical components in action potential propagation**

One of the key points in our understanding of the functioning of nervous systems in general was the discovery of the ionic mechanisms underlying the membrane potential and the action potential, powered by ionic pumping mechanisms. However, substantial, and overwhelming, evidence continues to accumulate to show that this is not the whole story, because the action potential is always accompanied by a synchronized coupled soliton pressure pulse in the neuronal cell membrane [5, 32] (Table 1), which taken together form the action potential pulse (APPulse) [5] which instigates channel opening (Fig. 2).

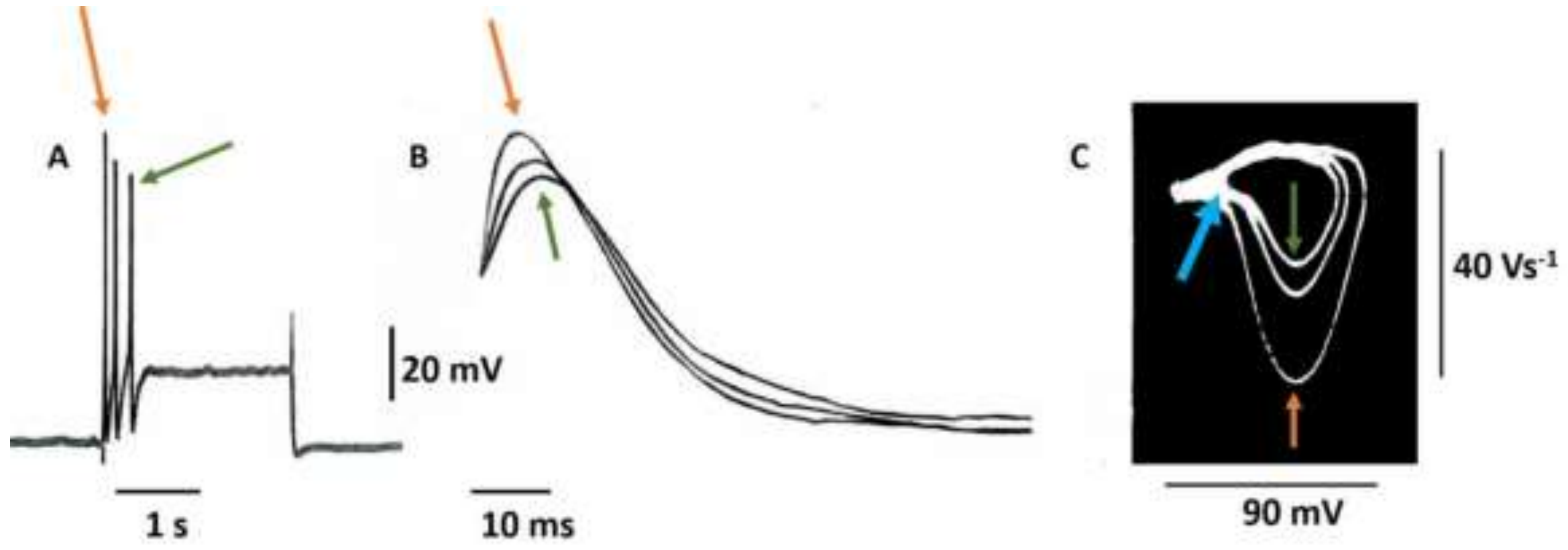

**Figure 1 – Plasticity of action potential shape and action potential peak** recorded from the soma of a fast-adapting pedal I cluster neuron (for details see Slade et al, 1981) in the intact brain of the mollusc Lymnaea stagnalis (L.). The cell was normally silent and activity was initiated by a 0.2nA current pulse of 3 s duration injected into the cell via a bridge balanced recording electrode. The same three spikes are represented in each case; a) on a slow time base, b) on a faster time base and c) as a phase plane portrait in which rate of change of voltage (dV/dt)is plotted against voltage itself and the inward depolarizing phase is displayed downward maintaining the voltage clamp convention. In each trace the peak of the first action potential is indicated by an orange arrow, the second action potential peak is unlabelled the third action potential peak is indicated by a green arrow. The three successive spike peaks clearly vary temporally from one another, but the threshold point of initiation remains constant as indicated in c) by the blue arrow in the phase plane portrait (From Winlow and Johnson [39,] licensed under Creative Commons BY-NC-SA 4.0).

- A 'soliton' mechanical pulse accompanies an action potential and is stable propagating at constant velocity [5,40,41]
- Electrically recorded action potentials are accompanied by optically detected movements of the action potential, which mimic the action potential [42,43]
- Non-linear sound waves/pressure pulses in lipid monolayers can show the main characteristics of nerve impulses [44,-47]
- Membrane oscillator theory suggests that ion currents associated with action potential cause vortex phenomena leading to pressure waves in the nerve cell membrane [48]
- Ion channel separation is too great to allow for ion channel interference from adjacent channels caused by ionic charge [49-52]
- Ion channels can be opened by mechanical stimulus [53-56]
- There is deformation of the membrane by activation of ion channels [4,40]
- Entropy (thermodynamic) measurements do not follow the H&H action potential but do follow the APPulse [57-61]
- These findings are supported by detailed mathematical modelling and computational simulations [62-69]

**Table 1 – The Action Potential Pulse (APPulse) – there is accumulating evidence for the soliton pressure pulse in neuronal membranes.**

Rather than detracting from the HH model, the biomechanical data should be seen to enhance it and should not be ignored.  Unless either component can be shown to be both necessary and sufficient to generate the action potential, we should assume that that membrane biophysics and biomechanics are inextricably linked to generate the action potential pulse.  In essence, Hodgkin and Huxley [3] demonstrated that sodium ions entering an axon generated the rise of the action potential and that this was counteracted by the opening of potassium gates to allow the outward flow of potassium ions. This was then thought to be followed by a decremental wave of depolarization spreading into the inactive region ahead of the action potential to cause further opening of sodium gates, thus allowing action potential propagation along the length of the axon.

Briefly, mechanical surface waves accompany action potential propagation [4,40,41] and these may be necessary to allow opening of ion channels ahead of the action potential since ion channel separation is too great to allow for spread of local current from one ion channel adjacent channels. [49-52] However, in the Hodgkin and Huxley model, action potential propagation depends on flow of charge from one channel to the next across the surface of the membrane, but we have demonstrated elsewhere that the distances involved are too great for this to occur it would be important for you to see it too [5] because the "channels are not crowded" together as shown by Hille in 1992. [104] Furthermore, there is good evidence that electrically recorded action potentials are accompanied by optically detected movements of the neuronal membrane [42,43] and also for deformation of the membrane by activation of ion channels. [4,40] What is more, ion channels can be opened by mechanical stimuli [53-56] as illustrated in Figure 2.

**MODULATION AND MEMORY STORAGE**

One of the major hurdles to overcome in our understanding of the nervous system is that of memory storage after learning. For example, bacteria such as *E coli* adapt to changing external conditions [70,71], have a memory of previous conditions and can modify their direction of locomotion accordingly via their "nanobrain". [72] Whether they are capable of associative learning and a form of cognition is open to speculation [71-73], but of course they do not rely on synapses to store their memory and it is possible that synapses are not the (sole) locus of memory stores in eukaryotes (see below).

Within nervous systems, many forms of chemical transmission occur, but of course chemical transmission introduces a time delay, much less so at electrical synapses. It is conjectured from comparative genomic and phylogenetic studies that the earliest transmitters were probably amines, peptides and gas transmitters such as nitric oxide [14,74,75], perhaps secreted into water flowing through the organisms.  In advanced bilaterians, there is a wide range of transmitters, often used in different ways, but fulfilling similar functions [75] and it has been suggested that that many transmitter pathways may have been transferred from bacteria to animals. [76]

Not all chemical synapses are equal, in that some have greater  synaptic weight and thus a greater influence than others on the follower cells to which they are connected, be those excitatory or inhibitory actions. Synapses also provide latency changes through the actions of neurotransmitters. Electrical synapses have reduced latency [77] compared with chemical synapses, but both together produce a spectrum of latencies depending upon their exact construction and location.  According to Katz, [78] "secretion of the transmitter is not synchronous with the arrival of the action potential in the nerve terminal, but lags well behind it" before even the first quantum of transmitter is released from a synaptic vesicle. [79] However, synaptic delay is modifiable, resulting in synaptic plasticity. [80] This can be related to changes in intracellular calcium concentration, which is important for transmitter release as demonstrated at neuromuscular junctions by Katz and Miledi. [81] Synaptic terminals can be affected by a variety of neuromodulatory agents which may be neurotransmitters, neurohormones or psychoactive substances [82] and can result in short-term plastic changes such as synaptic facilitation or depression and long-term potentiation or depression of synaptic events. Both the latency and weighting of synaptic responses may be affected [83] and these events can have profound consequences for the behaviour of the organism. [82]

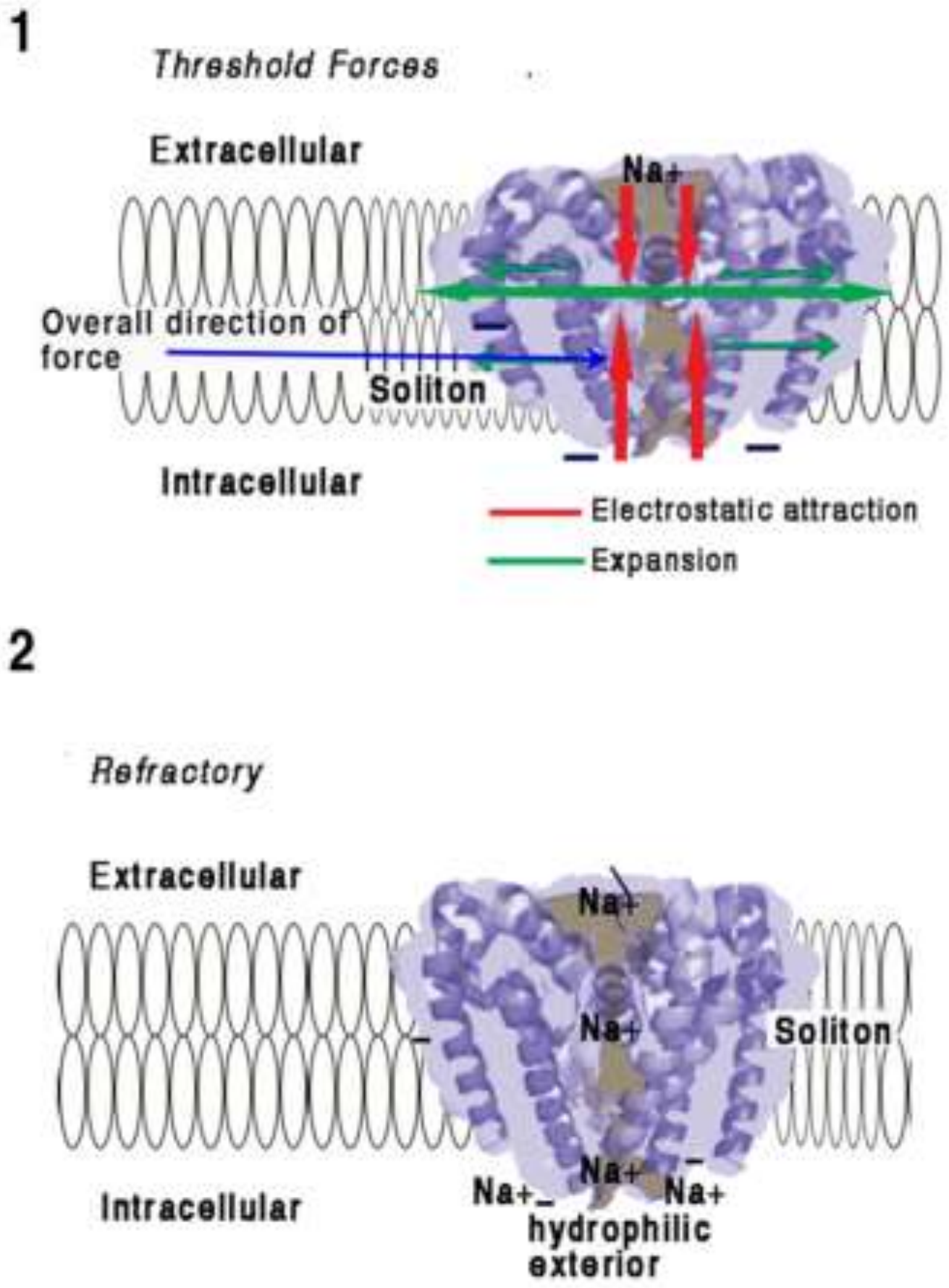

**Figure 2 – Instigation of channel opening by the APPulse.** 1. Pressure from the accompanying pressure wave of the action potential disturbs the ion channel electrostatic seal. Attracted electrostatically charged ions pass through the channel causing it to contract across the membrane. This in turn puts energy back into the pressure wave. 2. The ion channel becomes refractory when enough Na+ ions pass through to produce electrostatic equilibrium. Partially reconstructed from Johnson and Winlow [5] and McCusker et al [103] , used under Creative Commons BY-NC-SA 3.0

Currently, the role of synaptic plasticity in learning and memory is under intense scrutiny by cognitive scientists [84] with the suggestion that memories are molecularly stored within neurons and that plastic synaptic changes in weighting occur only after learning has occurred and been stored in memory. Support for this idea comes from work by on *Lymnaea stagnalis* [85,86] and on *Aplysia* [87] where it has been demonstrated that long term memory is stored in cell bodies. Thus, plastic changes in synaptic weighting may be "*a means of regulating behavior…only after learning has already occurred*". [84] Such assumptions would not be counter to findings that increased levels of associative learning efficiency are correlated with increased weighting of glutamatergic synapses and down-regulated by increased weighting of GABAergic inputs to neurons in the mouse barrel cortex. [88]

From the above it can be seen that the action potential is a plastic phenomenon and that major plastic changes can occur at nerve terminals. The question that needs to be posed is "how does the nervous system compute the information carried by and between nerve cells and over what sort of timescale?" Computers tend to work in nano or microseconds while nervous systems have been demonstrated to work on the basis of milliseconds. We will now attempt to resolve this conundrum.

**COMPUTATION WITHIN NERVOUS SYSTEMS**.

We have already described the orthodox physiological action potential as elucidated by Hodgkin and Huxley [3] and have suggested that it is accompanied by a soliton pressure wave, the action potential pulse. We suspect that these events are inextricably connected to one another and together with the computational action potential (CAP) (Figure 3) they form an ensemble of three inseparable concurrent states. At present it is not possible to determine which, if any, of these states predominates.

Most computational models of the action potential assume that it is a binary event, but it is most likely to be a quantum ternary event [33], the CAP. It moves along an axon at a defined speed determined by the transmission dynamics of the that particular membrane. As with the electrophysiological action potential the CAP has three well defined phases [32] :

- Phase 1 Resting potential of indeterminate length and retaining its integrity under normal resting conditions and actively maintained by membrane pumps
- Phase 2 The spike or digit resulting from ion changes and mechanical activity
- Phase 3 The refractory phase when no new action potentials may occur. Its duration varies according from one axon to another. Unlike the other two phases the refractory phase is an analogue variable, which is able to reroute action potentials along different pathways at bifurcations.

In computational terms action potential instigation occurs at threshold, not at the action potential peak as we demonstrated in Figure 1. Furthermore, phase-ternary computation occurs when collisions occur between coalescing action potentials across a membrane. [5,32] As a primary source of computation, it is *fast, accurate to microseconds, and efficient.*

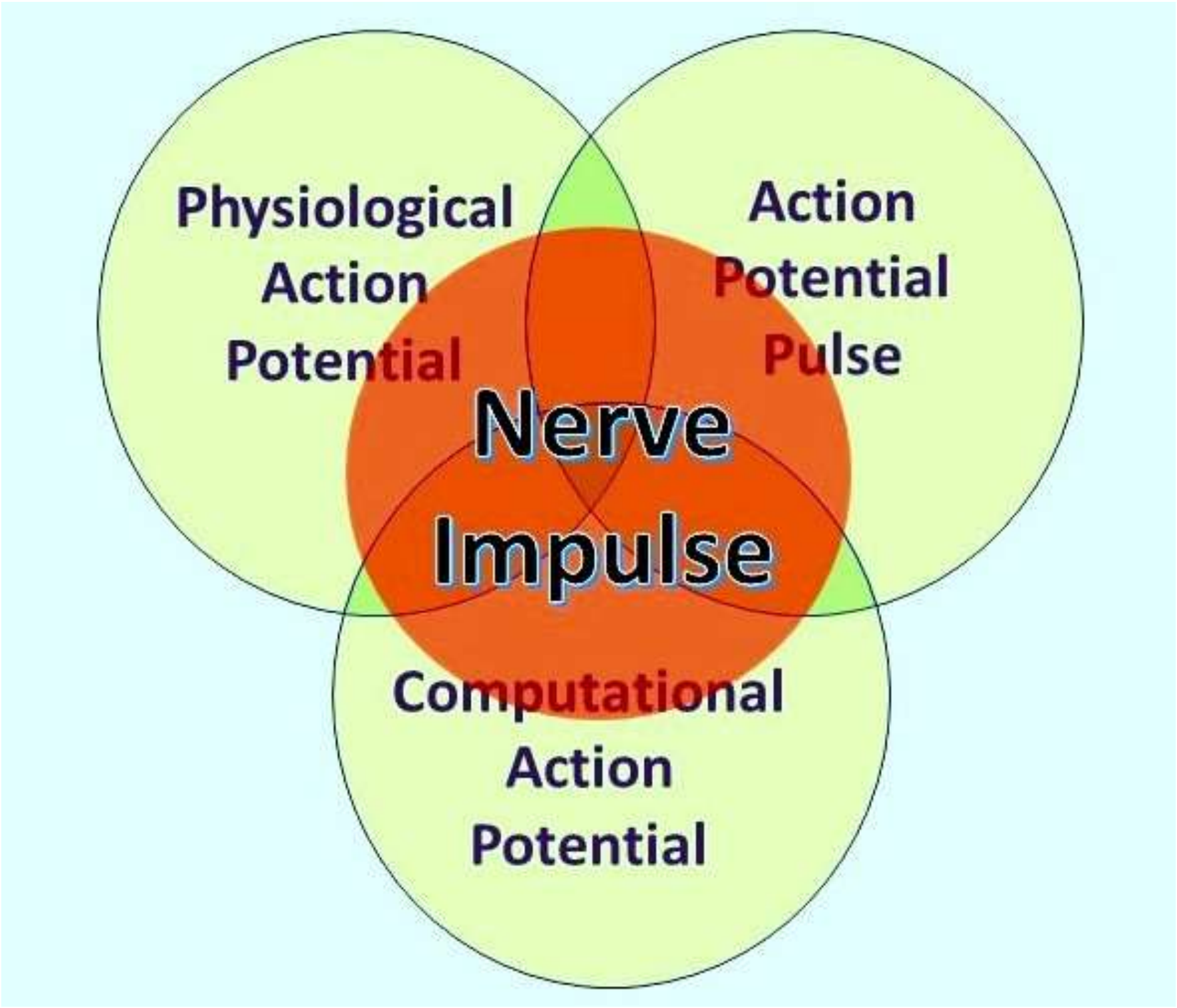

**Figure 3 – The nerve impulse may be and ensemble of three inseparable, concurrent states of the action potential.** What an observer will perceive depends on their investigational perspective. The physiological action potential is the orthodox action potential described in detail by Hodgkin and Huxley. [3] The action potential pulse is the mechanical pressure wave for which substantial evidence is presented in Table 1 and the computational action potential was first described by Johnson and Winlow. [32]

The basis of computation is that an outcome reflects inputs for any system. In figure 4A binary inputs are reflected in the outputs. If the binary inputs change the outputs will change accordingly. The process is not as simple for neurons which compute by differential frequencies. When two asynchronous CAPs collide the temporal component of the refractory period of the leading CAP will annul the second CAP. This is the same as when two opposing action potential cancel. CAP's are therefore quanta of ternary information computing temporally according to collisions. This temporal computation is important when considering the brain neural network where computation is not timed as in a conventional machine. As shown in figure 4B it permits temporal computation leading to changes in frequency across the network. In a previous paper we explored the shortcomings of artificial network models in neural computation. [89] Most models assume that processing in neural networks work like conventional binary computers, but in our view, this ignores the dynamic structure of real neural networks, operating by phase ternary computation. [89,90]

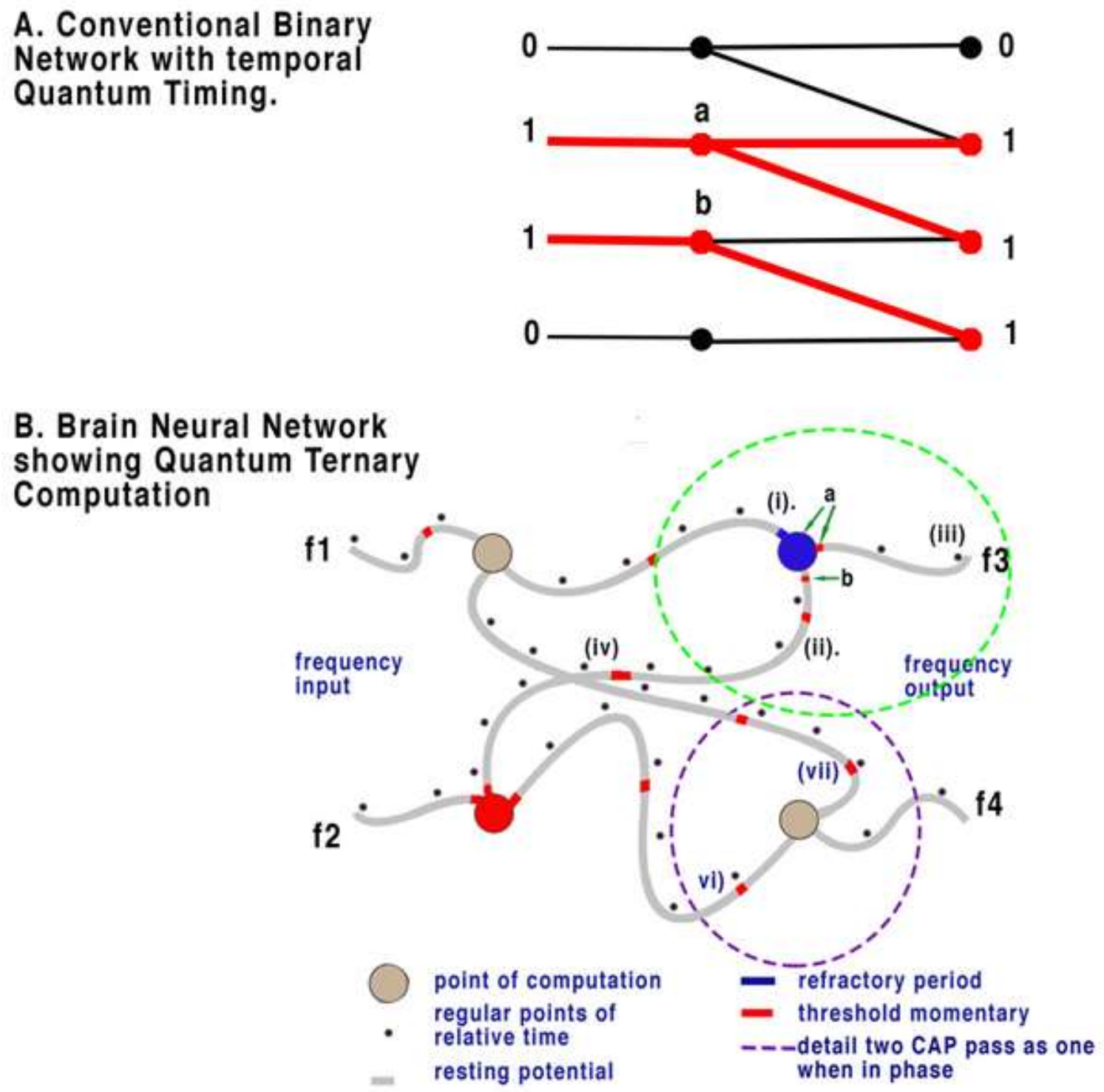

**Figure 4 – Computation within networks. A**. Illustration of conventional binary computing within a network of eight nodes. Inputs correspond to outputs logically. Nodes are gated by programming and the entire internodal activity is timed to synchronise. In this case at (a) there is a single quantum of information shown in red extending across the internodal distance. **B.** Represents a Brain Neural Network with four nodes, these are the convergences of neurons, synapses are part of internodal timing and are not shown. Timing is unrestricted and depends upon the structure of the neuron. Computational action potentials CAP travel along the surface of the membrane. Threshold is represented in red. CAP are not restricted to one per internodal length. In the green circle a convergence forms a node (i). (a) is a CAP represented by threshold(red) and refractory(blue). CAP (b) is timed to overlap with a and so will be annulled. The reduction and addition of CAP through the network changes the output frequencies of the output f3. Computation is by differential frequencies and not by base/clock timing. Therefore, changes in frequency of either f1 or f2 in this diagram result in corresponding frequency changes in f3 and f4.

## Quantum mechanics of computation and frequency.

The similarities of quantum events and computation between action potentials have recently been discussed [33] and all quantum computation in a network can be described as the result of a combination of collisions of quanta. In an un-programmed network, like the brain, this process is dependent upon the timing of each event. In a man-made computer, timing between nodes is defined by clock speed and the computation is binary. This results in the computation between nodes being synchronised from one node to the next (Figure 4A). Nodal transmission is therefore equivalent to one quantum. In a conventional quantum-computer the collisions of quanta are determined by the logic at each node predetermined by programming. In the brain neural network, action potentials pass over the neuron surface as quanta separated by frequency, so more than one quantum can exist within the internodal distance figure 4B.

The quanta in the brain can be visualised as action potentials, their propagation along a single neuron being binary with a temporal component. Collisions from action potential quanta reveal that the membrane at the point of collision has a refractory period during which no other quanta or action potentials can pass, resulting in their extinction. Recently [32,90] we demonstrated that as each action potential quantum collides at a convergence, the result computes in a phase-ternary manner, due to the membrane's refractory period. At the point of collision on the neuron membrane, each passing action potential exhibits a refractory period thus blocking subsequent action potentials for a specified time.

In the retina the activity of about 130 million light receptors (rods and cones) in each eye are coded down to around 1000,000 neurons in each optic nerve, a huge reduction from input to output which requires error free computation. [90,33] Consideration of the retinal 'circuit diagram' provided by Tsukamoto and Omi [91] strongly suggests that the retinal circuitry works by phase-ternary computation, rather than binary analysis. [90] In addition, the timing required for the digital period of computation must take place within 10 microseconds. Given that the timing of the peak of the action potential is temporally unstable (Figure 1), it is not possible that this is the source of computation. To compute two action potentials collision must be timed to interfere with 10 microsecond precision or less or computation cannot take place, [90] which implies that it cannot not be explained by a theoretical extrapolation into electrical cable theory.

As we demonstrated above (Figure 1) threshold is the most appropriate fixed point for neural computation. However, there is no clock in the brain or the retina (which is an extension of the brain) and in the absence of any timing computation cannot take place. In terms of computation by action potentials, there are four key aspects to consider: timing, accuracy, error redaction and speed of information transfer.

## Timing: There is no central clock in the brain

*Chronobiology* We are all aware of our daily circadian rhythms but these do not imply that there is central neural clock generating absolute times for our bodies. [92] This becomes obvious to us after a long airflight when these rhythms are disturbed and may take some days to recover, implying that they are entrained by the local environment and that there is no central clock in the brain. [32] Thus, time in the brain or neural network is not the same as the "absolute time" available from atomic clocks. However, in computational terms, modern computers are based around the concept of a Turing machine [93] and all contain a central system clock, unavailable in nervous systems, but there is a requirement for computational synchronisation within neural networks.

*Time and Neural Plasticity* Given that animal and human neural networks are able to accurately recall events over a lifetime of computation, the formation of memory must occur within a generally stable structure, implying that established structures exist for information processing and memory storage at times when plasticity is inoperative. Thus, computation in these real neural networks must be conducted, not according to contemporary linear time, but in accordance with the timing of plasticity (from milliseconds to years) whether short term, e.g. post-tetanic potentiation or long-term plasticity, e.g. due to adaptive stress. [94,95] Thus each computation would need to be processed when the network is within a stable state for a very short, but finite moment (in the order of microseconds – see below), before or after which the plastic change is made. Of course, the whole network does not have to be dynamically stable during this short period, only the section relevant to the processing of information that requires processing at that stable time location.

## Accuracy

Here we define accuracy in terms of the phase change as action potentials are deflected down different branches of multibranched neurons commonly found in nervous systems, as described elsewhere. [33,96,97] There are many forms of plasticity in neurons, but for accurate computation to take place, as in retinal processing ,[90] it is essential that temporal plasticity must be negligible in comparison to temporal timing. Contemporary models of neural networks assume the spike to be responsible for timing, but the model we present [32] indicates that threshold is responsible for timing the next action potential. Computation thus occurs between the threshold of one CAP and the refractory period of another at positions such as axonal branch points where action potentials can interact with or interfere with one another. Thus, the important attributes of the action potential are threshold, and temporal duration of its refractory period. [5] This implies that computation occurs by interference of one CAP with another often at specific points on the membrane resulting in either annulment of one action potential by another or the generation of trains of action potentials.

## Error redaction and noise reduction

Redaction of errors is tied up with accuracy of transmission and is important in both real and artificial brain neural networks. There is substantial noise generated in neural networks, [98] often due to spontaneous activity, but in parallel processing in axons leading to a single neuron such random events are cancelled out by collision with the refractory period of previous CAPs, thus reducing both error and noise in the system. [5,32]

## Speed of information transfer and Role of synapses

Synapses provide an important role in control of the neural network, but their purpose in computation is to set latencies and to alter the phase relationships of incoming action potentials. Synapses may also act as a slower parallel computational method than phase ternary computation to separate compartments within the neural network. However, phase ternary computation within a neural network is capable of instantaneously, but temporarily, storing information as fast working memory, regardless of any other memory storage or retrieval processes within that network. In the model proposed here, synapses are treated as nodal points in neural networks whose efficacy and phase can be varied. Chemical synapses typically transmit with variable latencies.

We conclude that, in a phase ternary network, memory is a function of associations between all previously recorded events and acts at the speed of the threshold. This is a previously unexplored memory concept. Each synchronisation of CAP requires a threshold of no more than 10 microseconds to digitise the computation. The variation in transmission time for an action potential is many times that. Each 10 microseconds a trit of information is carried temporally so that if two action potentials are carried the information is nonary (base9). This increases the memory capability of each neuron. Computation is by single action potentials in parallel in temporal space. It does not require immediate modification of synapses. However, synaptic weighting is modifiable in the aftermath of spike trains [99,100] and is important in learning and memory. [101,102] It is probable that Phase ternary computation within neural networks has general applicability.

In terms of general applicability to both nervous systems and Artificial Intelligence, our belief is that by using a model of the action potential that reflects its multiple modes of action we will be more able to model the mechanisms underlying functional neural networks, be they biological or computational.

## CONCLUSIONS

- Action potentials are evolutionarily ancient and may have arisen independently on several occasions. In nervous systems they subserve the functions of communication, modulation and computation
- It is possible that the some or all of these three functions apply to non-neuronal cells such as plant and bacterial cells and in carcinomas, but this hypothesis requires elucidation
- Orthodox Hodgkin-Huxley physiological action potentials appear to co-exist with, and may be inseparable from, the action potential pulse (soliton wave) and computational action potentials.
- A quantum model of computation explains how action potentials compute within a brain neural network such as the retina and by extension other parts of the brain.